\documentclass[letter,twocolumn]{jpsj2} 
\def\runauthor{Yutaka {\sc Shirata} {\it et al.}}

\title
{
Quantum Magnetization Plateau in Spin-1 Triangular-Lattice Antiferromagnet Ba$_3$NiSb$_2$O$_9$
}

\author
{ 
Yutaka \textsc{Shirata}, Hidekazu \textsc{Tanaka}\thanks{E-mail address: tanaka@lee.phys.titech.ac.jp}, Toshio \textsc{Ono}$^{1}$, Akira \textsc{Matsuo}$^{2}$, Koichi \textsc{Kindo}$^{2}$ and
Hiroki \textsc{Nakano}$^{3}$
}

\inst
{
Department of Physics, Tokyo Institute of Technology, Meguro-ku, Tokyo 152-8551\\
$^{1}$Department of Physical Science, Osaka Prefecture University, Sakai, Osaka 599-8531\\
$^2$Institute for Solid State Physics, University of Tokyo, Kashiwa, Chiba 277-8581\\
$^3$Graduate School of Material Science, University of Hyogo, Kamigori, Hyogo 678-1297
}

\recdate
{
\hspace{7cm}
}

\abst{We report the results of magnetization and specific heat measurements on Ba$_3$NiSb$_2$O$_9$, which is a quasi-two-dimensional spin-1 triangular-lattice antiferromagnet. We observed a nonclassical magnetization plateau at one-third of the saturation magnetization that is driven by spin frustration and quantum fluctuation. Exact diagonalization for a 21-site rhombic cluster was performed to analyze the magnetization process. Experimental and calculated results agree well.
}

\kword{Ba$_3$NiSb$_2$O$_9$, triangular-lattice antiferromagnet, magnetization plateau}

\begin{document}
\maketitle

Triangular-lattice antiferromagnet (TLAF), which is one of the simplest frustrated systems, exhibits a rich variety of physics\,\cite{Collins,Harrison,Balents}. A symbolic quantum effect predicted for small-spin Heisenberg TLAF is that magnetization is quantized at one-third of the saturation magnetization $M_{\rm s}$ in magnetic fields\,\cite{Nishimori,Chubukov,Nikuni,Honecker,Farnell}. For classical spin, the equilibrium condition in a magnetic field ${\mib H}$ is given by ${\mib S}_1\,{+}\,{\mib S}_2\,{+}\,{\mib S}_3\,{=}\,g{\mu}_{\rm B}{\mib H}/(3J)$ with the sublattice spins ${\mib S}_i$. Because the number of the equations  is smaller than the number of parameters to determine the spin configuration, the stable state is not uniquely determined; thus, the ground state is infinitely degenerate. This classical degeneracy can be lifted by the quantum fluctuation that is remarkable for small spin, and a specific spin state is selected as the ground state. Consequently, the {\it up-up-down} state is stabilized in a finite field range, which can be observed as a magnetization plateau at $\frac{1}{3}M_{\rm s}$\,\cite{comment_Nakano}.

In experiments, the quantum magnetization plateau at $\frac{1}{3}M_{\rm s}$ was actually observed in Cs$_2$CuBr$_4$\,\cite{Ono1,Ono2,Ono3,Fujii1,Fujii2,Tsujii}, where Cu$^{2+}$ ions having spin-$\frac{1}{2}$ form a spatially anisotropic triangular lattice with $J^{\prime}/J\,{=}\,0.74$. In Cs$_2$CuBr$_4$, an additional quantum plateau was also observed at $\frac{2}{3}M_{\rm s}$\,\cite{Ono2,Ono3}. Furthermore, precise thermodynamic measurements revealed the cascade of field-induced quantum phase transitions\,\cite{Fortune}. These unusual observations should be attributed to the spatially anisotropic triangular lattice and the Dzyaloshinsky-Moriya interaction\,\cite{Miyahara1,Alicea1,Alicea2,Tay,Starykh}, but the overall explanation is still an open question.

The motivation of this study comes from a question on whether the ground states for Heisenberg TLAFs with integer and half-integer spins are qualitatively the same. It is known that in the case of kagome-lattice Heisenberg antiferromagnet that is closely related to Heisenberg TLAF, the ground states for spin-$\frac{1}{2}$ and spin-1 are qualitatively different\,\cite{Hida}. For spin-1 Heisenberg TLAF, the ground states in zero and finite magnetic fields have not been well understood. NiGa$_2$S$_4$ is known as spin-1 Heisenberg TLAF with incommensurate spin correlation\,\cite{Nakatsuji,Takeya,Stock}. In this nickel compound, no clear magnetization plateau has been observed, although the magnetization shows a tiny undulate anomaly at around $\frac{1}{3}M_{\rm s}$\,\cite{Yamaguchi2}. To investigate the magnetization process of spin-1 Heisenberg TLAF, we have performed high-field magnetization measurements on Ba$_3$NiSb$_2$O$_9$. As shown below, a nonclassical magnetization plateau was observed at  $\frac{1}{3}M_{\rm s}$.

\begin{figure}[t]
\begin{center}
\includegraphics[width=5.6cm, clip]{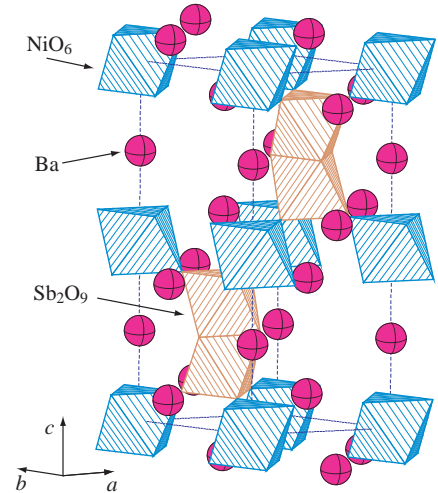}
\end{center}
\vspace{-2mm}
\caption{(Color) Crystal structure of Ba$_3$NiSb$_2$O$_9$. Dotted lines denote the chemical unit cell.}
 \label{fig:cryst}
\end{figure} 

Ba$_3$NiSb$_2$O$_9$ crystallizes in a high symmetric hexagonal structure, $P6_3/mmc$ \cite{Jacobson,Treiber}, which is the same as the hexagonal BaTiO$_3$ structure (see Fig.\,\ref{fig:cryst}). The structure is composed of single NiO$_6$ octahedra and face-connected Sb$_2$O$_9$ double octahedra, which are linked sharing corners. Magnetic Ni$^{2+}$ ions form layers of triangular lattice parallel to the $c$ plane, which are separated by the nonmagnetic layer of Sb$_2$O$_9$ double octahedra. Therefore, the interlayer exchange interaction is expected to be much smaller than the intralayer exchange interaction. Ba$_3$NiSb$_2$O$_9$ is considered to be one of the best candidates of  spin-1 Heisenberg TLAF. However, owing to the weak interlayer exchange interaction, Ba$_3$NiSb$_2$O$_9$ undergoes antiferromagnetic ordering at $T_{\rm N}\,{=}\,13.5$\,K\,\cite{Doi}. In this study, we also measured the specific heat to investigate the nature of the magnetic ordering.  As shown below, successive phase transitions were observed.

Ba$_3$NiSb$_2$O$_9$ powder was prepared via a chemical reaction 3BaCO$_3$ + NiO + Sb$_2$O$_5$ $\longrightarrow$ Ba$_3$NiSb$_2$O$_9$ + 3CO$_2$. Reagent-grade materials were mixed in stoichiometric quantities, and calcined at 1200 $^{\circ}$C for 30\,h in air. Ba$_3$NiSb$_2$O$_9$ was sintered at 1200$\,\sim$\,1600\,$^\circ$C for more than 30\,h after being pressed into a pellet. The sample obtained was examined by X-ray powder diffraction. A tiny peak due to nonmagnetic impurity BaSb$_2$O$_6$ of ${\sim}\,2$\,wt\% was observed.

The specific heat of Ba$_3$NiSb$_2$O$_9$ was measured down to 1.8\,K using a physical property measurement system (Quantum Design PPMS) by the relaxation method. High-field magnetization measurement of up to 53 T was performed using an induction method with a multilayer pulse magnet at the Institute for Solid State Physics, University of Tokyo. We also measured the temperature dependence of magnetic susceptibility and confirmed that the result was the same as that reported by Doi {\it et al.}\,\cite{Doi}.  

\begin{figure}[t]
\begin{center}
\includegraphics[width=7.5 cm, clip]{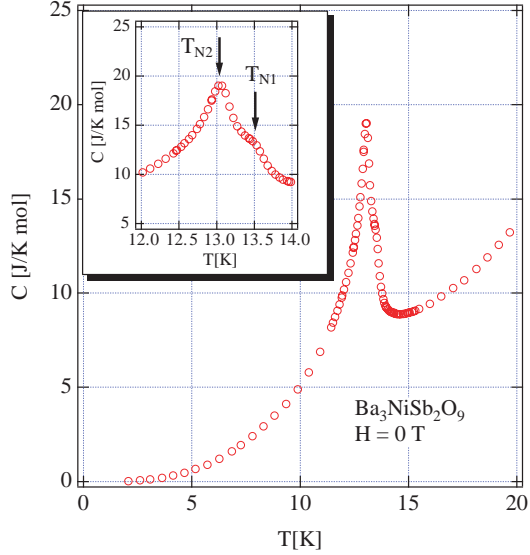}
\end{center}
\vspace{-2mm}
\caption{(Color) Specific heat of Ba$_3$NiSb$_2$O$_9$ measured at zero magnetic field. Inset shows an expansion at around 13\,K.}
 \label{fig:heat}
\end{figure}

Figure~\ref{fig:heat} shows low-temperature specific heat measured at zero-magnetic field. As shown in the inset of Fig.\,\ref{fig:heat}, Ba$_3$NiSb$_2$O$_9$ undergoes two magnetic phase transitions at $T_{\rm N1}\,{=}\,13.5$\,K and $T_{\rm N2}\,{=}\,13.0$\,K. In Heisenberg-like TLAF, the successive phase transition arises when the magnetic anisotropy is of easy-axis type, while a single transition occurs for the easy-plane anisotropy \cite{Matsubara}. The transition scenario for the easy-axis anisotropy is as follows: with decreasing temperature, the $c$ axis component of spins orders first at $T_{\rm N1}$ and then the $ab$ components order at $T_{\rm N2}$. Consequently, below $T_{\rm N2}$, spins form a triangular structure in a plane parallel to the $c$ axis. The reduced temperature range of the intermediate phase $(T_{\rm N1}-T_{\rm N2})/T_{\rm N1}$ is determined from the ratio of the easy-axis anisotropy to the exchange interaction \cite{Matsubara,Miyashita2}. The successive phase transitions with a very narrow intermediate phase in Ba$_3$NiSb$_2$O$_9$ indicate that the magnetic anisotropy is of the easy-axis type and is much smaller than the exchange interaction. The origin of the anisotropy should be a single-ion anisotropy of the form $D(S^z)^2$.

\begin{figure}[t]
\begin{center}
\includegraphics[width=8.3 cm, clip]{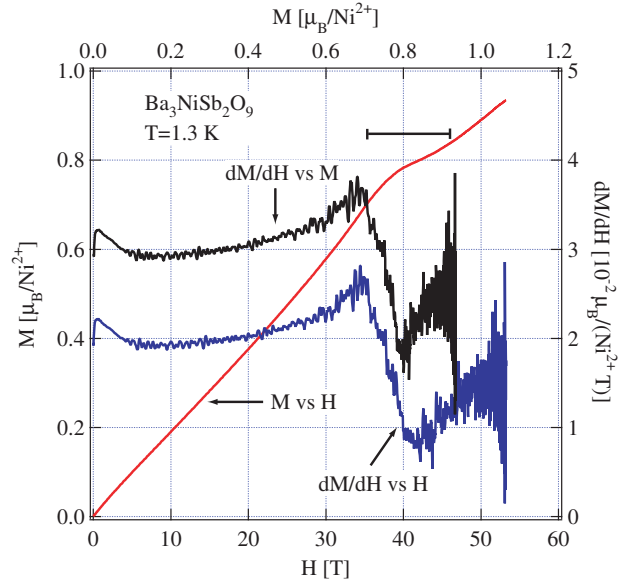}
\end{center}
\vspace{-2mm}
\caption{(Color) Magnetization curve, $dM/dH$ vs $H$ and $dM/dH$ vs $M$ of Ba$_3$NiSb$_2$O$_9$ measured at 1.3\,K. The data for $dM/dH$ vs $M$ is shifted upward by $1{\times}10^{-2} {\mu}_{\rm B}/({\rm Ni}^{2+}\,{\rm T})$. A horizontal bar denotes the field range of the magnetization plateau calculated by exact diagonalization.}
 \label{fig:MH}
\end{figure}

\begin{figure}[t]
\begin{center}
\includegraphics[width=7.6 cm, clip]{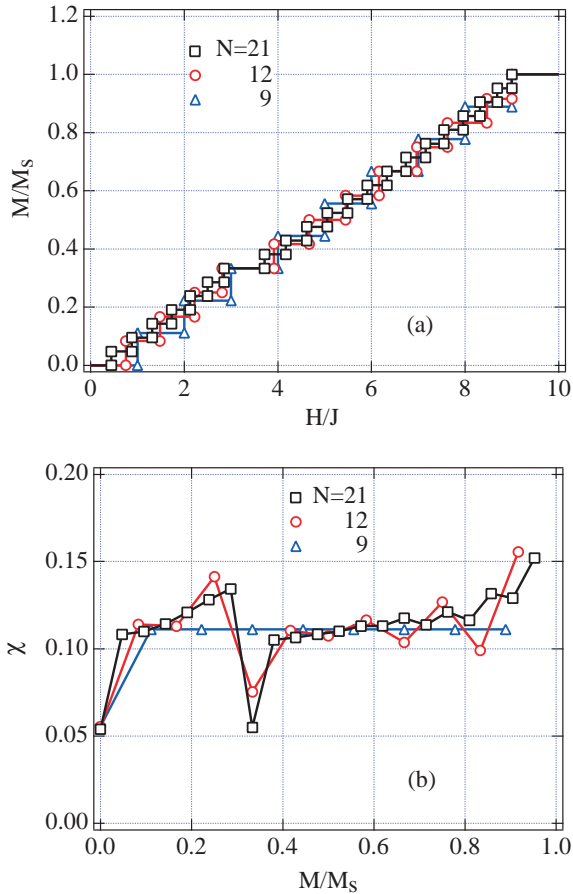}
\end{center}
\vspace{-2mm}
\caption{(Color) (a) Magnetization curves and (b) derivative susceptibility vs normalized magnetization $M/M_{\rm s}$ obtained by exact diagonalizations for 9-, 12-, and 21-site rhombic clusters.}
 \label{fig:MHcal}
\end{figure}

Figure~\ref{fig:MH} shows the magnetization curve of Ba$_3$NiSb$_2$O$_9$ powder measured at 1.3\,K and the derivative susceptibility $dM/dH$ vs magnetic field $H$. A magnetization plateau is observed at $M\,{\simeq}\,0.8\,{\mu}_{\rm B}$/Ni$^{2+}$. Because the $g$ factor of Ba$_3$NiSb$_2$O$_9$ powder is $g\,{\simeq}\,2.3$ \cite{Kohl}, the magnetization at the plateau corresponds to one-third of the saturation magnetization $M_{\rm s}$. The magnetization anomalies at the lower and higher edge fields are smeared. In general, the $g$ factors for $H\,{\parallel}\,c$ and $H\,{\perp}\,c$ are slightly different and the field ranges of the plateau for these field directions are also different owing to the weak anisotropy energy. These facts lead to the smearing of the edge fields of the plateau in a powder sample. 

In the classical Heisenberg TLAF, stable spin state is not uniquely determined in a magnetic field due to spin frustration. However, small easy-axis anisotropy and thermal and quantum fluctuations can stabilize a collinear up-up-down state in a finite field range. In the case of classical Heisenberg TLAF with the easy-axis anisotropy along the $c$ axis, the $\frac{1}{3}$-plateau appears for $H\,{\parallel}\,c$, while not for $H\,{\perp}\,c$ \cite{Miyashita,Kitazawa,Ishii}. Thus, for a powder sample of such classical system, $dM/dH$ at the plateau range should be about two-third of $dM/dH$ for $H$ smaller than the lower edge field. In the present system, the minimum of $dM/dH$ at the plateau range is smaller than half of $dM/dH$ for $H\,{<}\,30$\,T. This implies that the $\frac{1}{3}$-plateau appears irrespective of external field direction. In classical Heisenberg TLAF, thermal fluctuation stabilizes the up-up-down state at finite temperatures, so that a plateaulike magnetization anomaly occurs in the magnetization curve \cite{Kawamura}. The magnetic field range of the thermally stabilized collinear state decreases with decreasing temperature and becomes zero at $T\,{=}\,0$. The $\frac{1}{3}$-plateau in Ba$_3$NiSb$_2$O$_9$ has a magnetic field range of about 10\,T even at 1.3 K (${\ll}\,T_{\rm N2}$). Thus, we can deduce that the $\frac{1}{3}$-plateau in Ba$_3$NiSb$_2$O$_9$ arises from the quantum fluctuation\,\cite{Nishimori,Chubukov,Nikuni,Honecker,Farnell}. However, the magnetization at the plateau range is not completely flat. This should be ascribed to the increase in the sample temperature due to the magnetocaloric effect and the susceptibility of impurities. 

Following the method described by Troyer {\it et al.}~\cite{Troyer} for one-dimensional gapped excitations, we can derive the low-temperature molar susceptibility ${\chi}\,{=}\,N_{\rm A}n_{\rm m}g^2{\mu}_{\rm B}^2\exp (-{\beta}{\Delta})/(2{\pi}a)$, for two-dimensional gapped excitations with dispersion ${\epsilon}(k)\,{=}\,{\Delta}\,{+}\,ak^2$. Here, $n_{\rm m}$ is the number of gapped modes and $n_{\rm m}\,{=}\,2$ in the present case. When the external field is one-third of the saturation field $H_{\rm s}$, the coefficient $a$ is given by $a\,{=}\,(7/8)J$~\cite{Nikuni}. If the susceptibility $dM/dH{=}8{\times}10^{-3} {\mu}_{\rm B}/({\rm Ni}^{2+}\,{\rm T})$ at the center field of the $\frac{1}{3}$-plateau arises only from the finite temperature effect, the spin temperature is estimated to be $T\,{\simeq}\,2.1$\,K using ${\Delta}/k_{\rm B}\,{\simeq}\,4.5$\,K\cite{gap} and $J/k_{\rm B}\,{=}\,19$\,K, which is given below. 

Figure~\ref{fig:MHcal}(a) shows magnetization curve of spin-1 Heisenberg TLAF calculated by exact diagonalization for 9-, 12- and 21-site rhombic clusters \cite{comment_code}. The magnetization curve for spin-1 Heisenberg TLAF using exact diagonalization has already been reported by L\"{a}uchli {\it et al.}~\cite{Lauchli}. 
However, their calculations for $N\,{\geq}\,21$ are limited to a higher-field part of the magnetization. Note that our result for $N\,{=}\,21$ covers all the magnetization values. In Fig.~\ref{fig:MHcal}(b), we also present derivative susceptibility ${\chi}$ defined as 
\begin{eqnarray}
{\chi}^{-1}=\frac{E(N, M+1)-2E(N, M)+E(N, M-1)}{1/M_{\rm s}}\,,
\end{eqnarray}
where $E(N, M)$ is the lowest energy for a given $z$ component of the total spin $S_{\rm tot}^z\,{=}\,M$ 
for a given cluster size $N$. It is easier to detect the critical behavior around the plateau in $\chi$ rather than only in the magnetization curve~\cite{Nakano}. Our results in Figs.~\ref{fig:MHcal}(a) and (b) reveal a magnetization plateau at $\frac{1}{3}M_{\rm s}$, although the magnetic field range of the plateau is fairly small as compared with the spin-$\frac{1}{2}$ case \cite{Honecker,Farnell,Sakai}. 

As shown in Fig.~\ref{fig:MHcal}, the calculated susceptibility increases with increasing magnetization and shows a maximum just below $M\,{=}\,\frac{1}{3}M_{\rm s}$. Just at $M\,{=}\,\frac{1}{3}M_{\rm s}$, $\chi$ discontinuously becomes small. This behavior is also observed in $dM/dH$ vs $M$ shown in Fig. \ref{fig:MH}. For $M\,{>}\,\frac{1}{3}M_{\rm s}$, the calculated magnetization and susceptibility exhibit no anomaly suggestive of quantum phase transition up to saturation. 

The lower and higher edge fields tend to converge to $H_{\rm c1}\,{=}\,2.85J$ and $H_{\rm c2}\,{=}\,3.72J$, respectively, with increasing cluster size. A horizontal bar in Fig.~\ref{fig:MH} denotes the calculated field range of the $\frac{1}{3}$-plateau. Here, we chose the center field of the $\frac{1}{3}$-plateau as $(H_{\rm c1}\,{+}\,H_{\rm c2})/2\,{=}\,41$\,T. The experimental field range of the plateau is roughly estimated as ${\sim}\,60$\% of the calculated field range. From the center field of the plateau of 41\,T and $g\,{\simeq}\,2.3$, the exchange constant is evaluated as $J/k_{\rm B}\,{\simeq}\,19$\,K. This value of $J$ is approximately $\frac{3}{5}$ times as large as $J/k_{\rm B}\,{\simeq}\,32$\,K that is obtained from the Weiss constant ${\Theta}\,{=}\,{-}\,128$\,K and the molecular field theory.   

For the spin-1 case, the field range of the $\frac{1}{3}$-magnetization-plateau normalized by the saturation field is calculated as $(H_{\rm c2}\,{-}\,H_{\rm c1})/H_{\rm s}\,{=}\,0.097$, while for the spin-$\frac{1}{2}$ case, the normalized plateau range is 0.171 \cite{Farnell,Sakai}. The $\frac{1}{3}$-magnetization-plateau for the spin-1 case is considerably suppressed as compared with that for the the spin-$\frac{1}{2}$ case. The field range of the $\frac{1}{3}$-plateau and the $dM/dH$ vs $M$ observed in Ba$_3$NiSb$_2$O$_9$ are consistent with those calculated for the spin-1 Heisenberg TLAF. This indicates that the magnetization plateau observed in Ba$_3$NiSb$_2$O$_9$ arises from the quantum effect characteristic of the spin-1 Heisenberg TLAF. 

For TLAF with $S\,{\geq}\,1$, the magnetization plateau at $\frac{1}{3}M_{\rm s}$ can be stabilized with the help of the biquadratic exchange interaction of the form $-K_{ij}({\mib S}_i\,{\cdot}\,{\mib S}_j)^2$ with $K_{ij}\,{>}\,0$ \cite{Lauchli,Penc}, which mainly arises from the spin-lattice coupling as discussed by Penc {\it et al.} \cite{Penc}. The biquadratic exchange interaction may be responsible for the magnetization plateau observed for $H\,{\perp}\,c$ in RbFe(MoO$_4$)$_2$ \cite{Inami,Svistov1,Smirnov}, in which the normalized plateau range is as large as 0.10 despite spin-$\frac{5}{2}$. In Ba$_3$NiSb$_2$O$_9$, the contribution of the biquadratic exchange interaction should be minimal, because the field range of the magnetization plateau can be explained quantitatively within the bilinear exchange interaction.

As shown in Fig.~\ref{fig:MHcal}(b), the calculated susceptibility exhibits a sharp dip at $M\,{=}\,0$, almost irrespective of cluster size $N$. No such anomaly is observed for the spin-$\frac{1}{2}$ triangular- and kagome-lattice antiferromagnets without a spin gap \cite{Sakai,Nakano2}. At present, the origin of the susceptibility anomaly at $M\,{=}\,0$ is unclear. On the other hand, the susceptibility observed at small magnetization in Ba$_3$NiSb$_2$O$_9$ is finite because of magnetic ordering due to the weak interlayer exchange interaction.

In conclusion, we have presented the results of specific heat and high-field magnetization measurements on Ba$_3$NiSb$_2$O$_9$ that is described as a spin-1 Heisenberg TLAF. The present system exhibits two phase transitions $T_{\rm N1}\,{=}\,13.5$\,K and $T_{\rm N2}\,{=}\,13.0$\,K due to weak easy-axis anisotropy. We observed a magnetization plateau at $\frac{1}{3}M_{\rm s}$ at temperatures below $T_{\rm N2}$. We performed exact diagonalization for rhombic spin clusters with up to 21-sites to analyze the magnetization process. The calculated results are in agreement with experimental observations. The magnetization plateau observed in Ba$_3$NiSb$_2$O$_9$ arises from the interplay between spin frustration and the quantum effect characteristic of small-spin Heisenberg TLAF.

\section*{Acknowledgment}
This work was supported by a Grant-in-Aid for Scientific Research (Nos. 20244056, 20340096, 21740249 and 22014012) from the Japan Society for the Promotion of Science, and the Global COE Program ``Nanoscience and Quantum Physics'' at TIT funded by the Ministry of Education, Culture, Sports, Science and Technology of Japan. H.\,T. was supported by grant from the Mitsubishi Foundation. Part of the computations were performed using the facilities of the Supercomputer Center, Institute for Solid State Physics, University of Tokyo. 



\begin{thebibliography}{99} 

\bibitem{Collins} M. F. Collins and O. A. Petrenko: 
Can. J. Phys. \textbf{75} (1997) 605.
\bibitem{Harrison} 
A. Harrison: J. Phys.: Condens. Matter \textbf{16} (2004) S553.
\bibitem{Balents} L. Balents: Nature \textbf{464} (2010) 199.

\bibitem{Nishimori} H. Nishimori and S. Miyashita: 
J. Phys. Soc. Jpn. \textbf{55} (1986) 4448.
\bibitem{Chubukov} A. V. Chubukov and D. I. Golosov: 
J. Phys.: Condens. Matter \textbf{3} (1991) 69.
\bibitem{Nikuni} T. Nikuni and H. Shiba: 
J. Phys. Soc. Jpn. \textbf{62} (1993) 3268.
\bibitem{Honecker} A. Honecker: 
J. Phys.: Condens. Matter \textbf{11} (1999) 4697.

\bibitem{Farnell} D. J. J. Farnell, R. Zinke, J. Schulenburg and J. Richter: J. Phys.: Condens. Matter \textbf{21} (2009) 406002.

\bibitem{comment_Nakano} For spin-$\frac{1}{2}$ kagome-lattice antiferromagnet, exact diagonalization of up to 36 clusters showed that magnetization exhibits not the $\frac{1}{3}$-plateau but ramp anomaly. See ref.~\cite{Nakano}.

\bibitem{Nakano} H. Nakano and T. Sakai: J. Phys. Soc. Jpn. \textbf{79} (2010) 053707.

\bibitem{Ono1} T. Ono, H. Tanaka, H. Aruga Katori, F. Ishikawa, H. Mitamura and T. Goto: Phys. Rev. B \textbf{67} (2003) 104431.
\bibitem{Ono2} T. Ono, H. Tanaka, O. Kolomiyets, H. Mitamura, T. Goto, K. Nakajima, A. Oosawa, Y. Koike, K. Kakurai, J. Klenke, P. Smeibidle and M. Mei{\ss}ner: J. Phys.: Condens. Matter \textbf{16} (2004) S773. 
\bibitem{Ono3} T. Ono, H. Tanaka, T. Nakagomi, O. Kolomiyets, H. Mitamura, F. Ishikawa, T. Goto, K. Nakajima, A. Oosawa, Y. Koike, K. Kakurai, J. Klenke, P. Smeibidle, M. Mei{\ss}ner and H. Aruga Katori: J. Phys. Soc. Jpn. \textbf{74} (2005) Suppl. p. 135.
\bibitem{Fujii1} Y. Fujii, T. Nakamura, H. Kikuchi, M. Chiba, T. Goto, S. Matsubara, K. Kodama and M. Takigawa: Physica B \textbf{346-347} (2004) 45.
\bibitem{Fujii2} Y. Fujii, H. Hashimoto, Y. Yasuda, H. Kikuchi, M. Chiba, S. Matsubara and M. Takigawa: J. Phys.: Condens. Matter \textbf{19} (2007) 145237.
\bibitem{Tsujii} H. Tsujii, C. R. Rotundu, T. Ono, H. Tanaka, B. Andraka, K. Ingersent and Y. Takano: Phys. Rev. B \textbf{76} (2007) 060406(R).
\bibitem{Fortune} N. A. Fortune, S. T. Hannahs, Y. Yoshida, T. E. Sherline, T. Ono, H. Tanaka and Y. Takano: Phys. Rev. Lett. \textbf{102} (2009) 257201.

\bibitem{Miyahara1} S. Miyahara, K. Ogino and N. Furukawa: Physica B \textbf{378-380} (2006) 587.
\bibitem{Alicea1} J. Alicea and M. P. A. Fisher: Phys. Rev. B \textbf{75} (2007) 144411.
\bibitem{Alicea2} J. Alicea, A. V. Chubukov and O. A. Starykh: Phys. Rev. Lett. \textbf{102} (2009) 137201. 
\bibitem{Tay} T. Tay and O. I. Motrunich: Phys. Rev. B \textbf{81} (2010) 165116.
\bibitem{Starykh} O. A. Starykh, H. Katsura and L. Balents: Phys. Rev. B \textbf{82} (2010) 014421.

\bibitem{Hida} K. Hida: J. Phys. Soc. Jpn. \textbf{69} (2000) 4003.

\bibitem{Nakatsuji} S. Nakatsuji, Y. Nambu, H. Tonomura, O. Sakai, S. Jonas, C. Broholm, H. Tsunetsugu, Y. Qiu and Y. Maeno: Science \textbf{309} (2005) 1697. 
\bibitem{Takeya} H. Takeya, K. Ishida, K. Kitagawa, Y. Ihara, K. Onuma, Y. Maeno, Y. Nambu, S. Nakatsuji, D. E. MacLaughlin, A. Koda and R. Kadono: Phys. Rev. B \textbf{77} (2008) 054429.
\bibitem{Stock} C. Stock, S. Jonas, C. Broholm, S. Nakatsuji, Y. Nambu, K. Onuma, Y. Maeno and J.-H. Chung: Phys. Rev. Lett. \textbf{105} (2010) 037402.
\bibitem{Yamaguchi2} H. Yamaguchi, S. Kimura, M. Hagiwara, Y. Nambu, S. Nakatsuji, Y. Maeno, A. Matsuo and K. Kindo: J. Phys. Soc. Jpn. \textbf{79} (2010) 054710.

\bibitem{Jacobson} A. J. Jacobson and A. J. Calvert: J. Inorg. Nucl. Chem. \textbf{40} (1978) 447.
\bibitem{Treiber} U. Treiber and S. Kemmler-Sack: Z. Anorg. Allg. Chem. \textbf{487} (1982) 161.
\bibitem{Doi} Y. Doi, Y. Hinatsu and K. Ohoyama: J. Phys.: Condens. Matter \textbf{16} (2004) 8923.

\bibitem{Matsubara} F. Matsubara: J. Phys. Soc. Jpn. \textbf{51} (1982) 2424.

\bibitem{Miyashita2} S. Miyashita and H. Kawamura: J. Phys. Soc. Jpn. \textbf{54} (1985) 3385.

\bibitem{Kohl} P. K\"{o}hl and D. Reinen: Z. Anorg. Allg. Chem. \textbf{433} (1977) 81.

\bibitem{Miyashita} S. Miyashita: J. Phys. Soc. Jpn. \textbf{55} (1986) 3605.
\bibitem{Kitazawa} H. Kitazawa, H. Suzuki, H. Abe, J. Tang and G. Kido: Physica B \textbf{259-261} (1999) 890.
\bibitem{Ishii} R. Ishii, S. Tanaka, K. Onuma, Y. Nambu, M. Tokunaga, T. Sakakibara, N. Kawashima,
Y. Maeno, C. Broholm, D. P. Gautreaux, J. Y. Chan and S. Nakatsuji: Eur. Phys. Lett. \textbf{94} (2011) 17001.

\bibitem{Kawamura} H. Kawamura and S. Miyashita: J. Phys. Soc. Jpn. \textbf{54} (1985) 4530.

\bibitem{Troyer} M. Troyer, H. Tsunetsugu and D. W\"{u}rtz: Phys. Rev. B \textbf{50} (1994) 13515.
\bibitem{gap} The gap $\Delta$ at the center of the plateau was estimated from ${\Delta}=(1/2)g{\mu}_{\rm B}{\Delta}H$ assuming that the field range of the plateau ${\Delta}H$ is 6\,T for the powder sample.

\bibitem{comment_code} The program code was originally developed in the study of precise estimation of the Haldane gap in ref.~\cite{Nakano_Terai}. 
\bibitem{Nakano_Terai} H. Nakano and A. Terai: J. Phys. Soc. Jpn. \textbf{78} (2009) 014003.

\bibitem{Lauchli} A. L\"{a}uchli, F. Mila and K. Penc: Phys. Rev. Lett. \textbf{97} (2006) 087205.

\bibitem{Sakai} T. Sakai and H. Nakano: Phys. Rev. B \textbf{83} (2011) 100405(R).

\bibitem{Penc} K. Penc, N. Shannon and H. Shiba: Phys. Rev. Lett. \textbf{93} (2004) 197203.

\bibitem{Inami} T. Inami, Y. Ajiro and T. Goto: J. Phys. Soc. Jpn. \textbf{65} (1996) 2374.
\bibitem{Svistov1} L. E. Svistov, A. I. Smirnov, L. A. Prozorova, O. A. Petrenko, L. N. Demianets and A. Ya. Shapiro: Phys. Rev. B \textbf{67} (2003) 094434.
\bibitem{Smirnov} A. I. Smirnov, H. Yashiro, S. Kimura, M. Hagiwara, Y. Narumi, K. Kindo, A. Kikkawa, K. Katsumata, A. Ya. Shapiro and L. N. Demianets: Phys. Rev. B \textbf{75} (2007) 134412.

\bibitem{Nakano2} H. Nakano and T. Sakai: J. Phys. Soc. Jpn. \textbf{80} (2011) 053704


\end{thebibliography}
\end{document}